# Controlling magnetization reversal in Co/Pt nanostructures with perpendicular anisotropy

M. Tofizur Rahman,[1] Randy K. Dumas,[2] Nasim Eibagi,[2] Nazmun N Shams,[1] Yun-Chung Wu,[1] Kai Liu[2,*] and Chih-Huang Lai[1,*]

[1]*Department of Materials Science and Engineering, National Tsing Hua University, Taiwan*

[2]*Physics Department, University of California, Davis, California 95616, USA*

**Abstract**

We demonstrate a simple method to tailor the magnetization reversal mechanisms of Co/Pt multilayers by depositing them onto large area nanoporous anodized alumina (AAO) with various aspect ratios, $A$ = pore depth/diameter. Magnetization reversal of composite (Co/Pt)/AAO films with large $A$ is governed by strong domain-wall pinning which gradually transforms into a rotation-dominated reversal for samples with smaller $A$, as investigated by a first-order reversal curve method in conjunction with analysis of the angular dependent switching fields. The change of the magnetization reversal mode is attributed to topographical changes induced by the aspect ratio of the AAO templates.

Patterned magnetic nanostructures with perpendicular anisotropy have potential applications in a host of emerging technologies, such as bit patterned media,[1, 2] percolated perpendicular media,[3] spintronics,[4] magnetic random access memory[5] and magnetic sensors.[6] To realize the full potential of these technologies there are still key issues to be addressed: 1) simple and cost-effective fabrication process to achieve magnetic nanostructures over large areas and 2) understanding and control of magnetization reversal process in elements whose sizes are approaching the ferromagnetic exchange length. Among the many efforts to fabricate patterned magnetic nanostructures, a simple way is to deposit materials on pre-patterned substrates[1-3, 7] that, e.g., have been prepared by electron-beam lithography,[1, 2] nano-imprint lithography,[8] extreme ultraviolet lithography,[7] and di-block copolymer templates.[9, 10] In particular, anodized alumina (AAO) templates[3, 11] have been proposed for high-throughput large area substrate patterning, with potential of achieving sub-10 nm pattern sizes. On the other hand, magnetization reversal mechanisms of thin films with perpendicular anisotropy are known to be sensitive to film microstructures[12, 13] as well as nanopatterning.[1, 14] In nanostructures with length scales approaching the domain-wall width and exchange length, it is especially interesting to tailor the reversal mechanisms through structural engineering.

In this Letter, we report on the controlled modification of magnetization reversal mechanisms of Co/Pt multilayers (MLs) by depositing them on nanoporous AAO templates with different aspect ratios. The reversal mechanisms are studied by both the angular dependence of the switching fields and a first-order reversal curve (FORC) technique.

Hexagonal close-packed arrays of nanopores with a density of $\sim 4\times 10^{10}/cm^2$ are formed by anodic oxidization of a 50nm Al layer sputtered onto Si substrates in 0.3 M sulfuric acid under a constant voltage of 25 V.[3, 15] The pore diameter $D$, inter-pore edge-to-edge distance $d$, and aspect ratio $A$ (pore depth / pore diameter) are tuned to the desired values by chemical etching of fabricated AAO/Si in 5 wt% phosphoric acid at 30°C for different time, $t$. With increasing etching time, $D$ increases whereas $d$ and the height of the porous alumina film decrease, leading to an overall decrease in $A$. The average surface roughness of the contiguous network is virtually identical (0.4 - 0.5 nm) for all the templates.[16] Multilayers of Pt (3nm)/[Co (0.5nm)/Pt (2nm)]$_5$ are then deposited onto the various AAO/Si templates by DC magnetron sputtering at room temperature. We refer to these composite structures as (Co/Pt)/AAO/Si hereafter. For comparison, the same film structure is also deposited directly on a Si wafer. Fig. 1 shows the top-view scanning electron microscopy (SEM) images of the (Co/Pt)/AAO/Si films with different aspect ratios and the insets show the corresponding cross-sectional transmission electron microscopy (TEM) images. For samples with $A \geq 1.6$, Co/Pt films are deposited primarily on top of the AAO membrane and around the perimeter of the pores. However, for $A \sim 0.7$, significant amount of Co/Pt is deposited inside the pores.

Magnetic properties are measured by an alternating gradient and vibrating sample magnetometer (AGM/VSM). The continuous Co/Pt and composite (Co/Pt)/AAO/Si films all exhibit strong perpendicular anisotropy with magnetic easy axis along the film normal. In addition, angular dependence of the remanent coercivity is studied where the angle $\theta$ between the applied field and easy axis is varied [Fig. 2(a)]. The remanent coercivity or switching field $H_s$ is defined as the reverse field required to reduce the magnetization to zero. In order to uncover the fundamental interactions present, the $\Delta M$ method[17, 18] is utilized to obtain $\Delta M(H) = M_{DCD}(H)/M_R + 2M_{IRM}(H)/M_R - 1$, where $M_{DCD}(H)$, $M_{IRM}(H)$, and $M_R$ are the DC demagnetization, isothermal, and saturation remanence, respectively. $\Delta M(H)$ is expected to be positive for interactions dominated by exchange and negative for interactions with dipolar origins. Furthermore, the FORC technique is employed to capture ~$10^2$ reversal curves, each starting at successively more negative reversal fields $H_R$ and measured with increasing applied field $H$ perpendicular to the film plane.[13, 19] Families of FORC's for the continuous film, $A$~ 3.2, and $A$~0.7 samples are shown in Fig. 3 left column. A mixed second order derivative of the magnetization $M(H, H_R)$ is used to generate the FORC distribution[13, 20] $\rho \equiv -\partial^2 M(H, H_R)/2\partial H \partial H_R$ which is then shown in a contour plot (Fig. 3 right column). The FORC distribution can also be represented in ($H_C$, $H_B$) coordinates, where $H_C$ is the local coercive field and $H_B$ is the interaction field, through a rotation of the coordinate system defined by: $H_B=(H+H_R)/2$ and $H_C=(H-H_R)/2$.

We first note that the continuous film shows a monotonic and dramatic increase in $H_s$ with $\theta$ [Fig. 2(a)]. This is a clear sign that the reversal is dominated by domain-wall motion with negligible pinning.[3] The FORC distribution of this sample confirms this behavior. Because the mixed second order derivative eliminates purely reversible components of the reversal,[20] regions of irreversible switching can be easily determined by features in the FORC diagrams. Indeed the FORC distribution of the continuous Co/Pt film [Fig. 3(b)] is similar to earlier studies using FORC and transmission x-ray microscopy.[13] The two highly irreversible regions directly correspond to the nucleation and annihilation of reversed domains within the film. At $H_R$ = -200 Oe, (along line scan 1) there is a pronounced horizontal ridge which corresponds to the rapid nucleation and propagation of reversed domains in the film. The second prominent feature is the negative—positive vertical pair of peaks along $H_R$ = -600 Oe (line scan 2) which indicate the subsequent annihilation of domains as the sample approaches negative saturation. In between line scans 1 and 2 there is relatively little feature in the FORC distribution and the value of $\rho$ is very small, due to mostly reversible switching at -200 Oe > $H_R$ > -600 Oe. As seen previously in similar films, at this $H_R$ field range the width of the reversed domains expand while those unreversed domains contract and the overall topology of the domain patterns remains largely unchanged.[13] As the continuous film has not been deliberately patterned to induce artificial defects, it has little to impede domain wall motion and therefore has an extended region of reversible switching.

When the (Co/Pt) MLs are deposited on AAO/Si, the coercivity dramatically increases from 230 Oe in the continuous film to 1370 Oe for $A \sim 7.0$ and 1500 Oe for $A \sim 1.6$. This is due to the increase of effective pinning by the pores[21, 22] as pore diameter increases with decreasing aspect ratio. Additionally, the angular variation of $H_s$ shows a qualitatively different behavior as compared to the continuous film, Fig. 2(a). For $\theta < 45°$ the $A \sim 7.0$ and 3.2 samples show little variation in $H_s$, which is due to the pinning effect induced by the pores as confirmed by micromagnetic simulations.[3] As shown in Fig. 1, in the $A \sim 7.0$ and 3.2 samples most of the Co/Pt is on top of the AAO and in a contiguous network. Therefore, pinning of domain walls is dominant in those samples. The FORC features of the $A \sim 3.2$ sample [Fig. 3 (d)] are qualitatively very similar to the continuous film sample [Fig. 3 (b)], but with some significant differences. The horizontal ridge along line scan 1 identifying domain nucleation and propagation has now moved to $H_R = -1430$ Oe, as a result of the coercivity enhancement. The negative — positive pair of peaks, which indicate domain annihilation along line scan 2 ($H_R = -1790$ Oe), have become harder to distinguish as they overlap significantly with the nucleation peak along line scan 1. The overlap region in between line scans 1 and 2 demonstrates significant irreversible switching. In fact this FORC distribution is characteristic of polycrystalline hard/soft exchange spring magnets:[23] the magnetically hard and soft phases are respectively the Co/Pt deposited on top of the AAO and those near the perimeters of the pores; the two phases are physically connected, giving rise to the exchange coupling. This is also consistent with the positive peak found in the $\Delta M$ plot shown in Fig. 2(b) that indicates strong exchange interactions. The FORC distribution, along with the enhanced coercivity, clearly indicates the role of the pores as local pinning sites that impede the otherwise reversible motion of domain growth. The FORC analysis of the $A\sim7$ sample yields similar conclusions.

As the aspect ratio is reduced further, a fundamentally different reversal mechanism begins to emerge as more material is deposited into the pores. For the $A\sim1.6$ and 0.7 samples, $H_s$ now develops a pronounced minimum near 45° [Fig. 2(a)], which is a clear indication of a Stoner-Wohlfarth (S-W) like rotation-dominated reversal. For reference, the expected S-W behavior is shown as a solid line in Fig. 2(a). As $A$ decreases, the pore size increases and the edge-to-edge distance $d$ is reduced. For the $A\sim0.7$ sample the average $d$ (also the network width) is about 18-20 nm, comparable to the domain wall width of Co/Pt MLs (~15nm).[24] The Co/Pt on top of the AAO is in a contiguous network which reverses the magnetization primarily by rotation.[9] As shown in Fig.1 for the $A\sim1.6$ and 0.7 samples a significant amount of material is deposited into the pores. $\Delta M$ measurements show a strong negative peak for the A~1.6 and 0.7 samples [Fig. 2(b)], indicating strong dipolar interactions. This result suggests that the materials within the pores are physically separated from those on top of the AAO. The isolated material in the pores tends to reverse by rotation given their small volume. Further evidence of a rotation dominated reversal is provided by the FORC analysis. Fig. 3(f) shows the FORC distribution in the ($H_C$, $H_B$) coordinate system for the $A=0.7$ sample, which is characteristic of reversal via single domain rotation of the magnetization in nanodots[19] and nanoparticles.[25] As expected, the single peak is centered at the coercivity of the major loop (~1450 Oe) and has a distribution along the $H_C$-axis, indicating a distribution of switching fields in the sample.

In summary, we have demonstrated a simple method to tune the switching field and magnetization reversal mechanisms of Co/Pt MLs with perpendicular anisotropy. As expected, the continuous film sample reverses by the relatively unimpeded motion of domain walls. However, the composite (Co/Pt)/AAO/Si films show altered reversal mechanisms depending on the aspect ratio of the AAO template they are deposited on. For high aspect ratios ($A \sim 7.0$ and 3.2) the Co/Pt primarily lies on top of the AAO and the reversal is dominated by the highly pinned motion of domain walls. As the aspect ratio is decreased to 1.6 and 0.7, magnetization reversal is primarily by rotation as the lateral dimensions of Co/Pt deposited both on top of AAO and inside the pores approach the domain wall width. These composite films also demonstrate an attractive fabrication procedure to achieve patterned media over large areas.

The work at NTHU is supported by the National Science Council of Republic of China under Grant No. NSC 95-2221-E-007-063-MY2, NSC 95-2112-M-007-054-MY3 and NSC 97-2622-E-007-002 respectively. Work at UCD is supported by CITRIS and U. C. Davis.

# References

\* Corresponding authors. E-mail: (C.H. L.) chlai@mx.nthu.edu.tw, (K.L.) kailiu@ucdavis.edu.

[1] T. Thomson, G. Hu, and B. D. Terris, Phys. Rev. Lett. **96**, 257204 (2006).

[2] O. Hellwig, A. Moser, E. Dobisz, Z. Z. Bandic, H. Yang, D. S. Kercher, J. D. Risner-Jamtgaard, D. Yaney, and E. E. Fullerton, Appl. Phys. Lett. **93**, 192501 (2008).

[3] M. T. Rahman, N. N. Shams, Y. C. Wu, C. H. Lai, and D. Suess, Appl. Phys. Lett. **91**, 132505 (2007).

[4] S. Mangin, D. Ravelosona, J. A. Katine, M. J. Carey, B. D. Terris, and E. E. Fullerton, Nature Materials **5**, 210 (2006).

[5] S. Assefa, J. Nowak, J. Z. Sun, E. O'Sullivan, S. Kanakasabapathy, and W. J. Gallagher, J. Appl. Phys. **102**, 063901 (2007).

[6] S. Khizroev, Y. Hijazi, R. Chomko, S. Mukherjee, R. Chantrell, X. Wu, R. Carley, and D. Litvinov, Appl. Phys. Lett. **86**, 042502 (2005).

[7] F. Luo, L. J. Heyderman, H. H. Solak, T. Thomson, and M. E. Best, Appl. Phys. Lett. **92**, 102505 (2008).

[8] G. M. McClelland, M. W. Hart, C. T. Rettner, M. E. Best, K. R. Carter, and B. D. Terris, Appl. Phys. Lett. **81**, 1483 (2002).

[9] K. Liu, S. M. Baker, M. Tuominen, T. P. Russell, and I. K. Schuller, Phys. Rev. B **63**, 060403 (2001).

[10] I. Bita, J. K. W. Yang, Y. S. Jung, C. A. Ross, E. L. Thomas, and K. K. Berggren, Science **321**, 939 (2008).

[11] M. T. Rahman, N. N. Shams, and C. H. Lai, Nanotechnology **19**, 325302 (2008).

[12] S. B. Choe and S. C. Shin, Phys. Rev. B **57**, 1085 (1998).

**Figure Captions**

**Fig.1.** SEM top-view and TEM cross-sectional images (inset) of Co/Pt MLs deposited on AAO/Si with different aspect ratios. The corresponding etching time $t$ (sec), pore diameter $D$ (nm), pore depth $h$ (nm) and interpore edge-to-edge distance $d$ (nm) are also shown.

**Fig. 2.** (a)Switching field, $H_s$ with different aspect ratio $A$ as a function of the angle $\theta$ between the applied field and the easy axis. (b) $\Delta M$ plots for two representative samples of $A \sim 3.2$ and 0.7.

**Fig. 3.** Families of FORC's, whose starting points are represented by black dots, are shown in (a), (c), (e) for the continuous film, $A\sim3.2$, and 0.7 samples respectively. The corresponding FORC distributions are shown in (b), (d), (f).

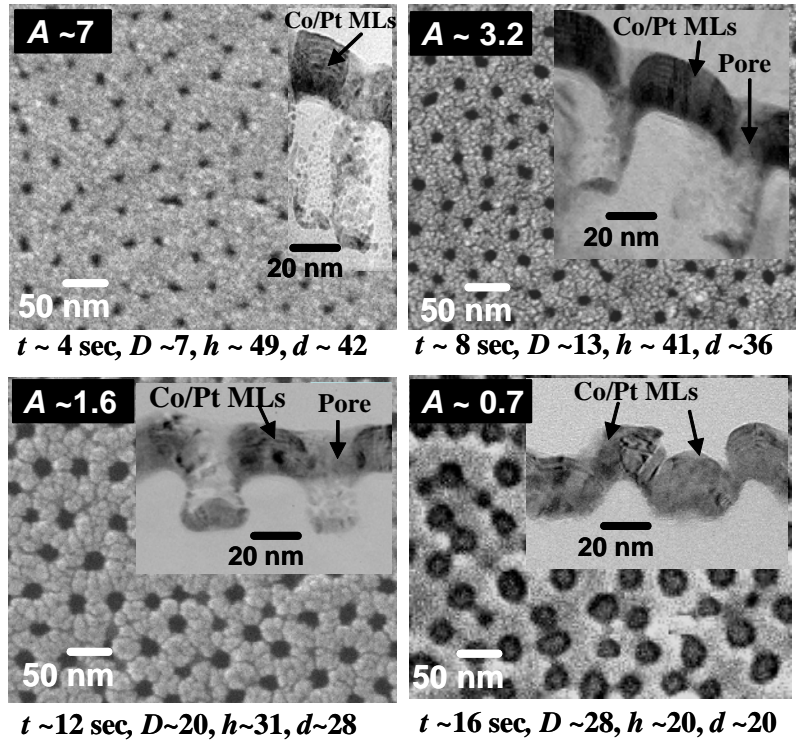

*t* ~ 4 sec, *D* ~7, *h* ~ 49, *d* ~ 42    *t* ~ 8 sec, *D* ~13, *h* ~ 41, *d* ~36

*t* ~12 sec, *D*~20, *h*~31, *d*~28    *t* ~16 sec, *D* ~28, *h* ~20, *d* ~20

**Fig. 1. Rahman et al.**

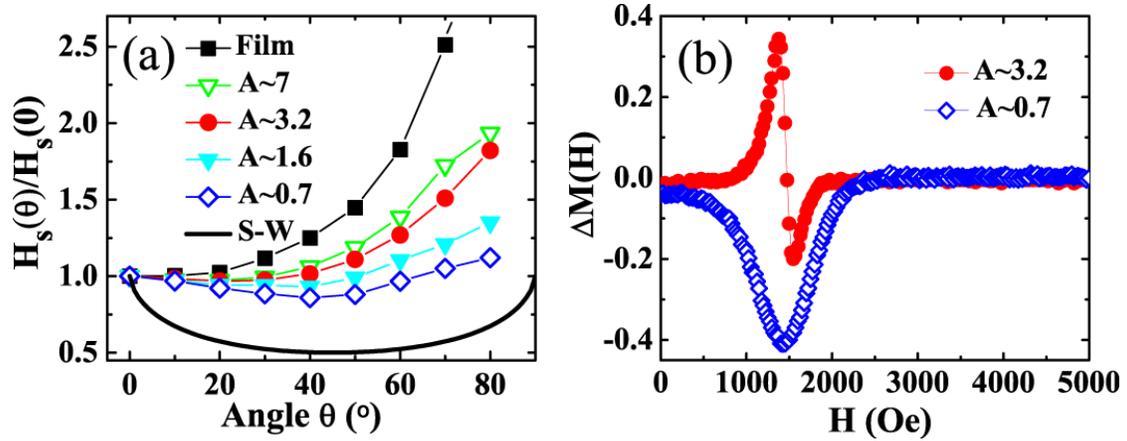

**Fig. 2. Rahman et al.**

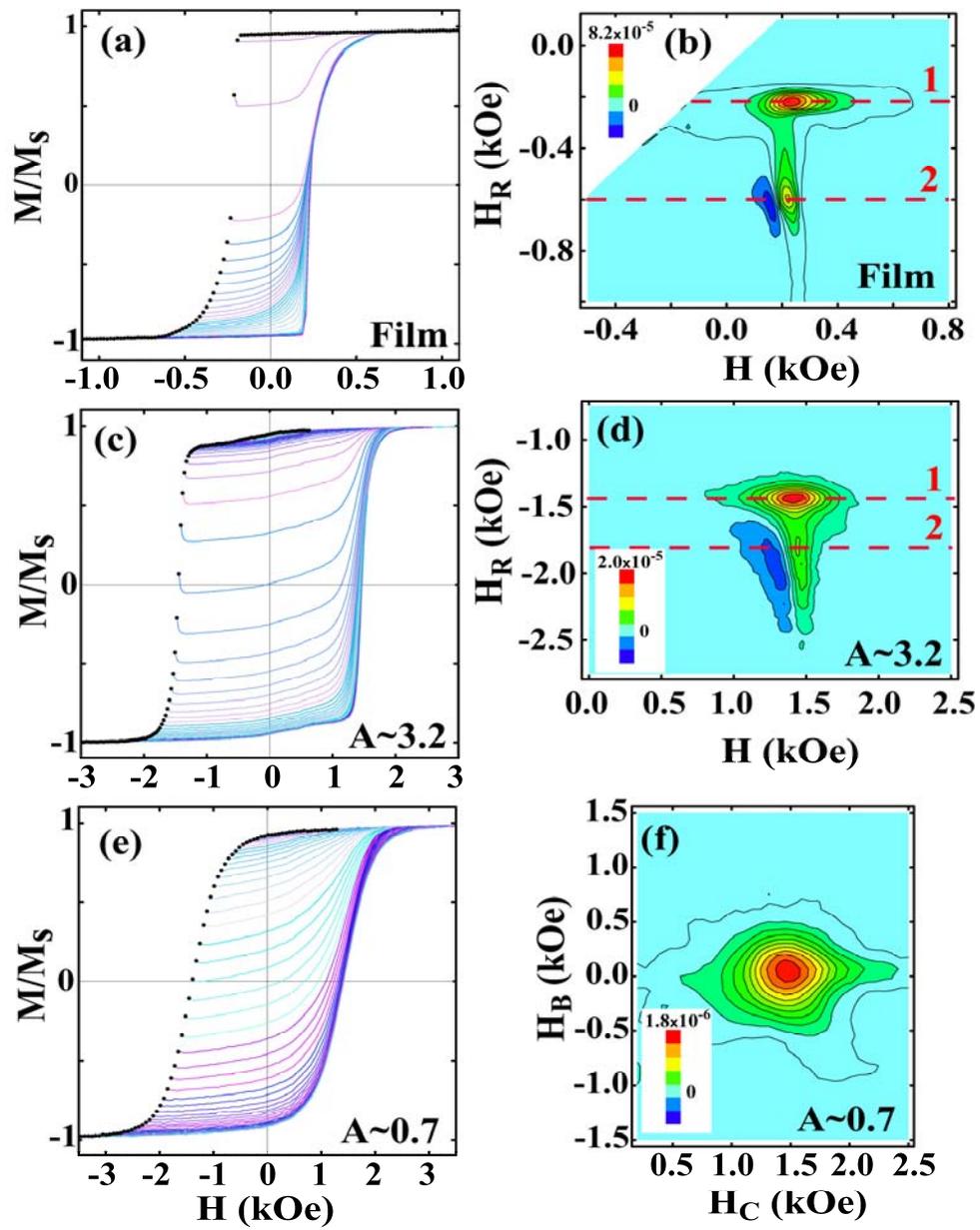

Fig. 3. Rahman et al.